\documentclass{aa}  

\usepackage{graphicx}
\usepackage{txfonts}
\begin{document} 

   \title{Towards a more complete sample of binary central stars of planetary nebulae with \textit{Gaia}}

  \author{N. Chornay\inst{1} \and N. A. Walton\inst{1} \and D. Jones\inst{2,3} \and H.M.J. Boffin\inst{4} \and M. Rejkuba\inst{4} \and R. Wesson\inst{5}}

  \institute{Institute of Astronomy, University of Cambridge, Madingley Road, Cambridge CB3 0HA, UK\\
              \email{njc89@ast.cam.ac.uk}
    \and
    Instituto de Astrof\'isica de Canarias, E-38205 La Laguna, Tenerife, Spain
              \and
              Departamento de Astrof\'isica, Universidad de La Laguna, E-38206 La Laguna, Tenerife, Spain
    \and
    ESO, Karl-Schwarzschild-str. 2, 85748 Garching, Germany
    \and
    Department of Physics and Astronomy, University College London, Gower Street, London WC1E 6BT, UK}

   \date{February 2021}

% \abstract{}{}{}{}{} 
% 5 {} token are mandatory
 
  \abstract
  % context heading (optional)
  % {} leave it empty if necessary  
   {Many if not most planetary nebulae (PNe) are now thought to be the outcome of binary evolutionary scenarios. However only a few percent of PNe in the Milky Way are known to host binary systems. The high precision repeated observing and long time baseline of \textit{Gaia} make it well suited to detect new close binaries through photometric variability.}
  % aims heading (mandatory)
   {We aim to find new close binary central stars of PNe (CSPNe) using data from the \textit{Gaia} mission, building towards a statistically significant sample of post common envelope, close binary CSPNe.}
  % methods heading (mandatory)
   {As the vast majority of \textit{Gaia} sources do not have published epoch photometry, we use the uncertainty in the mean photometry as a proxy for determining the variability of our CSPN sample in the second \textit{Gaia} data release. We derive a quantity that expresses the significance of the variability, and consider what is necessary to build a clean sample of genuine variable sources.}
  % results heading (mandatory)
   {Our selection recovers a large fraction of the known close binary CSPN population, while other CSPNe lying in the same region of the parameter space likely represent low-hanging fruit for ground-based confirmatory followup observations. \textit{Gaia} epoch photometry for four of the newly identified variable sources confirms that the variability is genuine and consistent with binarity.}
  % conclusions heading (optional), leave it empty if necessary 
   {}

   \keywords{binaries: close -- planetary nebulae: general -- Methods: statistical}

   \maketitle
%
%-------------------------------------------------------------------

\section{Introduction}

Planetary Nebulae (PNe) are the product of low and intermediate mass stars that underwent extensive mass loss during the Asymptotic Giant Branch (AGB) phase. The textbook version of the story is that PNe are the swan song of stars between 1 and 8 solar masses that are nearing the ends of their lives. Such stars shed their outer layers, growing brighter and hotter and ionising the newly formed nebula before ultimately cooling and fading into white dwarfs. While this explanation may indeed hold for some PNe, it has become increasingly clear in recent decades that many of these stars do not die alone: a significant fraction, if not a majority, of PNe are produced by systems that have undergone binary interactions \citep{miszalski2009oglebinaries,boffinjones2019book}. Particularly compelling evidence comes from the diverse range of morphologies observed in PNe \citep{balick+frank2002}, the most dramatic of which are thought to come from the presence of a close binary companion \citep{miszalski2009b,jonesboffinnaturebinaries2017}. Some authors have even suggested that binarity might be a requirement for the formation of a PN, which would resolve the tension between population synthesis models and the unexpectedly low number of PNe observed in the Milky Way \citep{demarcobinaries2006}.

Close binary central stars of PNe are not only relevant for the study of PNe: they also place important constraints on common-envelope (CE) theory \citep{iben+livio1993, yungelson1993}, a poorly understood evolutionary phase that is critical for the formation of a wide variety of astrophysical phenomena, including cataclysmic variables, low-mass x-ray binaries, type Ia supernovae, mergers, and gravitational wave sources \citep{jones2020}. A statistically significant sample of close binary systems is crucial for observationally constraining the orbital period distribution, which, in turn, is an important probe of the CE efficiency \citep{toonen2013}.

Such a sample is lacking: the identification of close binary central stars has been a painstaking process, with only 15 such systems revealed in the 30 or so years following the discovery of the first example \citep[see Chapter 3 of][for a historical overview]{boffinjones2019book} and the current total standing around 70. Binarity manifests itself in various ways. Widely separated pairs can be detected visibly, particularly from space \citep{ciardullo1999hstbinaries, benetti2003hstbinaries, liebert2013hstbinaries}, and bright but close companions produce blended spectra \citep[which can sometimes be confirmed by long-term, high-resolution radial velocity monitoring;][]{jones2017b}. Radial velocity variations are the most conclusive evidence but are often difficult to measure due to contamination from bright nebula lines and spurious variability stemming from post-AGB winds \citep{demarco2004}. The most dramatic photometric variability comes from eclipses \citep[requiring a lucky alignment of the orbital plane, e.g.;][]{jones2019} and/or the irradiation of a cool companion \citep[e.g.;][]{munday2020ethos1}. In very close systems, tidal distortion can result in the presence of ellipsoidal modulation in the observed light curve \citep[e.g.;][]{santandergarcia2015}. Detection of photometric variations is also subverted by the presence of a bright nebula and aggravated by variable seeing conditions for ground-based monitoring \citep{jones2015}. Space-based photometric monitoring has shown promising results, with variability in PN central stars having recently been identified in Kepler/K2 \citep{demarco2015kepler, jacoby2020kepler} and TESS \citep{aller2020tess}, though the high sensitivity of these missions is somewhat offset by the low spatial resolution, which dilutes the signal, particularly in the presence of nearby stars or bright nebulosity.

The European Space Agency's \textit{Gaia} mission \citep{gaiamission} offers great promise as it tracks not only positional but also photometric variations, and does so for the whole sky, including most known PN central stars \citep{chornay2020cspn}. \textit{Gaia} offers high spatial resolution and long time baselines at the cost of irregular sampling and a long wait for the availability of full epoch photometry for all sources. This paper illustrates the potential of the \textit{Gaia} mission for the discovery and characterisation of new binary PN central stars.

\section{Variability from mean photometry} \label{sect:methods}

Over the course of \textit{Gaia}'s mission, any given source is observed many times. The pattern of the observations is dictated by \textit{Gaia}'s scanning law \citep{gaiamission}, and the frequency at which a given source is observed depends on its position in the sky. The scanning law is such that a source should be observed at least six separate times over the course of a year, though due to \textit{Gaia}'s two telescopes and its rate of rotation, each of those so-called visibility periods generally has at least two observations of the source, separated by a few hours \citep{eyer2017variability}.

A crossing of a source across \textit{Gaia}'s focal plane is called a field-of-view (FoV) transit. During an FoV transit a source's flux is measured multiple times across different CCDs and through different passbands. \textit{Gaia} has three passbands: the main, wide $G$ band, and the blue $G_\textup{BP}$ and red $G_\textup{RP}$ bands. In the $G$ band there are typically eight or nine CCD transits per FoV transit; in  $G_\textup{BP}$ and $G_\textup{RP}$ there is one at most.

For most sources, \textit{Gaia}'s intermediate data releases include only the mean photometry in each band, that is, the mean flux $\overline{I_s}$ and its uncertainty $\sigma_{\overline{I_s}}$. In the calculation of these quantities individual CCD flux measurements are weighted by their nominal precisions \citep[Eq. 6 and Eq. 7 in][]{carrasco2016gaiadr1photomtery}. However the uncertainty in the mean is based on the observed measurement scatter, so it may not match that expected from the uncertainties of the individual fluxes. In particular, if a source is variable, this will manifest as increased scatter and a higher uncertainty. If variability is the dominant source of the scatter, the amplitude of that variability can be derived from the published uncertainty.

\subsection{Variability amplitude}

The uncertainty in the mean flux of a source published in the  \textit{Gaia} catalogue is the weighted standard deviation of the individual flux measurements divided by the square root of the number of CCD transits $N_{CCD}$ (corresponding to the \texttt{phot\_g\_n\_obs} column). The weights are not published, but if we assume that the weights are equal, we can recover the standard deviation of the fluxes in the $G$ band as $\sqrt{N_{CCD}}\ \sigma_{\overline{I_G}}$, where now $\overline{I_G}$ and $\sigma_{\overline{I_G}}$ are the mean and uncertainty for the $G$ band flux.

Following \citet{mowlavi2020lavs}, \citet{belokurov2017}, and others, we adopt the variability statistic
\begin{equation} \label{eq:a_varG}
    A_{var,G} \equiv \sqrt{N_{CCD}}\ \frac{\sigma_{\overline{I_G}}}{\overline{I_G}},
\end{equation}
which is simply the standard deviation of the individual CCD fluxes as a proportion of the mean flux.
This is related to the scatter in the magnitude as follows. Recall that the \textit{Gaia} $G$ band magnitude $G = -2.5 \log_{10} \overline{I_G}\ +\ G_0$, where $G_0$ is the zeropoint.
For a scatter in the individual flux measurements that is small relative to the mean flux ($A_{var,G}\ll1$), the standard deviation $\sigma_m$ of the corresponding magnitudes is
\begin{equation} \label{eq:sigma_m}
    \sigma_m = \frac{2.5}{\ln{10}}\ A_{var,G},
\end{equation}
using the properties of the derivative (i.e. the first-order Taylor polynomial).

An even more convenient relation comes from assuming that the variation is sinusoidal, as it often approximately is for close binary systems. In that case, $\sigma_m = A\ / \sqrt2$, where $A$ is the sinusoid's semi-amplitude. Equivalently, the range 2$A$ is then
\begin{equation} \label{eq:a_sine}
\textup{range}\ G = 2\sqrt{2}\ \sigma_m \approx 3.07\ A_{var,G}
\end{equation}
where $\sigma_m$ is as in Eq. \ref{eq:sigma_m}. For the full derivation and a deeper discussion see \citet{mowlavi2020lavs}.

\subsection{Variability significance}

The intrinsic measurement uncertainty varies significantly with magnitude, increasing for fainter sources due to photon statistics as well as including noticeable features at other magnitudes related to differences in data acquisition and calibration. It is important to ascertain whether the apparent variability that we obtain is actually significant. If we had the per-FoV transit uncertainties, we could determine an excess noise statistic analogous to the unit weight error for astrometric measurements. However, we do not have those, and, moreover, the errors are heteroscedastic, depending not only on source colour and magnitude, but varying temporally and spatially due to factors such as straylight conditions and sky background. Nonetheless the considerations are broadly similar to those discussed in \citet{gaiaruwe} in the context of \textit{Gaia}'s re-normalised unit-weight error (RUWE) statistic for astrometry.

With the aim of obtaining a crude significance statistic, we suppose that there is an intrinsic uncertainty in the flux measurement for a single FoV transit that depends on a source's magnitude and colour, $\sigma_\textup{G} = \sigma_\textup{G}(G,C)$ (where colour is the magnitude difference between the $G_\textup{BP}$ and $G_\textup{RP}$ bands, $G_\textup{BP}$\ --\ $G_\textup{RP}$). We refer to the standard deviation of the observed fluxes across different FoV transits for a given source using $\sigma_\textup{G, obs}$; in general this will only match $\sigma_\textup{G}$ in the limit of a large number of samples.

The arrangement of CCDs on \textit{Gaia}'s focal plane results in sources having an average of 8.86 $G$ band flux measurements per FoV transit \citep{jordigaiaphotometry}. As CCD transits occur sequentially and the FoV transit time is tens of seconds \citep{riello2018dr2photometricprocessing}, intrinsic source variability on any much larger timescale contributes to flux scatter only across different FoV transits. Thus we focus on FoV rather than CCD transits, and define the effective number of samples $k \equiv N_{FOV} \equiv N_{CCD}/8.86$.

For constant sources, values of $\sigma_\textup{G, obs}^2$ will follow a $\chi^2$ distribution with $k-1$ degrees of freedom:
\begin{equation}
    k\ \sigma_\textup{G, obs}^2 \sim \sigma_\textup{G}^2\ \chi^2_{k-1}.
\end{equation}
This will not match the distribution that we observe, not only because of the simplifying assumptions, but because all stars are variable at some level. However the mode of the distribution should not be too affected by the addition of these intrinsically variable sources, so, as the mode of $\chi^2_{k-1}$ is $(k-1)-2$, we can find $\sigma_\textup{G}$ as
\begin{equation}
    \sigma_\textup{G}^2 = \textup{mode} \left(\frac{k\ \sigma_\textup{G, obs}^2}{(k-1)-2} \right),
\end{equation}
assuming that the empirical distribution of $\sigma_\textup{G, obs}^2$ over which the mode is taken is appropriately smoothed.

For sufficiently large $k$, the $\chi^2_{k-1}$ distribution can be approximated as a normal distribution with matching mean $k-1$ and variance $2(k-1)$:
\begin{equation}
    \frac{k\ \sigma_\textup{G, obs}^2}{\sigma_\textup{G}^2} \sim \chi^2_{k-1} \approx \mathcal{N}(k-1,2(k-1)).
\end{equation}
This can be related to a standard normal by subtracting the mean and dividing by the square root of the variance:
\begin{equation}
    \sqrt{\frac{k-1}{2}} \left( \frac{k\ \sigma_\textup{G, obs}^2}{(k-1)\ \sigma_\textup{G}^2} - 1 \right) \sim \mathcal{N}(0,1).
\end{equation}
The term on the left is a dimensionless measure of significance, which can be further approximated for convenience as a natural logarithm using the relation $x\approx\ln{(1+x)}$ for small $x$ (as the part inside the parenthesis is close to zero). Furthermore, an additional magnitude- and colour-dependent scale factor can be added to ensure that the variance is indeed 1 (in practice, to ensure that the left side of the distribution has variance 1, as the right side includes most variable sources). This gives
\begin{equation} \label{eq:D}
    D \equiv \frac{1}{D_0} \sqrt{\frac{k-1}{2}} \ln{\left( \frac{k\ \sigma_\textup{G, obs}^2}{(k-1)\ \sigma_\textup{G}^2} \right)}
\end{equation}
where $D$ is the variability significance and $D_0 = D_0(G,C)$ is the scale factor. It is straightforward to compare this form of the expression to the logarithm of our variability amplitude $A_{var,G}$.

In Eq. \ref{eq:D} the significance of any observed variability increases with the number of FoV transits $k$. However as $k$ increases, the simplifying assumptions (rather than statistical properties) will increasingly contribute to the observed scatter.

We determine $\sigma_G$ and $D_0$ empirically using a representative sample of \textit{Gaia} sources. We bin sources by magnitude and colour and calculate $\sigma_G$ and $D_0$ using the distribution of flux uncertainties in each bin. Interpolating between these values yields two-dimensional functions of magnitude and colour that we use to compute the variability significance for individual sources.

\section{Results}

We apply our methodology to the CSPN catalogue of \citet{chornay2020cspn}. That catalogue was generated using an automated method to match CSPNe with sources in \textit{Gaia} data release 2 (DR2), and includes a reliability parameter (between 0 and 1) that indicates the certainty of the match. We expect that \textit{Gaia}'s early data release 3 (EDR3) will not significantly change the results; in particular it does not contain any new epoch photometry or variability analysis that can be used to validate our assumptions, and any new detections will be for faint sources where only the most extreme variability would be detectable. We join our catalogue with a compilation of known and suspected binary CSPNe from literature,\footnote{\url{http://www.drdjones.net/bcspn/}, as of October 18, 2020.} excluding one extragalactic PN and six objects identified as non-PNe in the Hong Kong/AAO/Strasbourg H$\alpha$ (HASH) PN catalogue of \citet{hashpn}, while adding PHR J1040-5417 \citep{hillwig2017phrj1040}.

\subsection{Known binaries}

\begin{figure}
    \centering
    \includegraphics[width=\hsize]{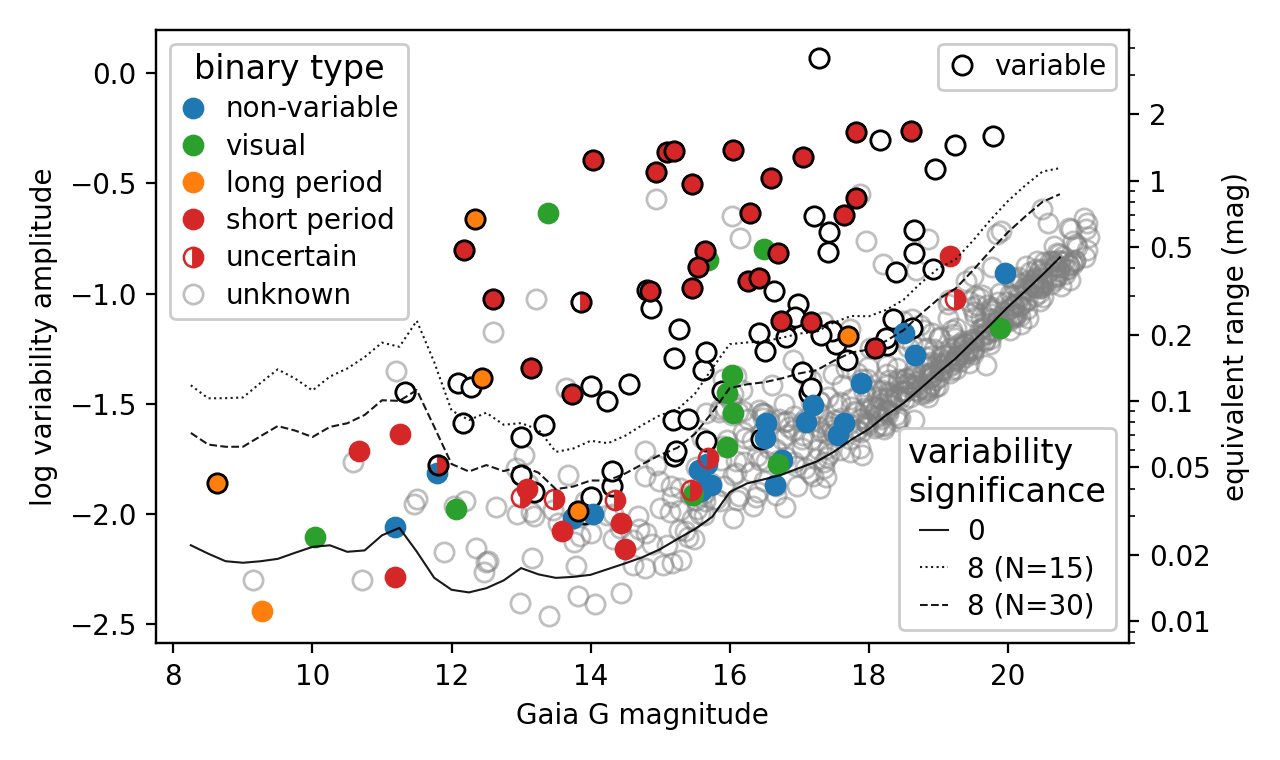}
    \caption{Variability amplitude versus \textit{Gaia} $G$ magnitude for CSPNe in our sample. Filled markers are known or suspected binary systems of different observational types; see text for description. Lines indicate constant variability significance $D$ for different numbers of FoV transits $N$.
    }
    \label{fig:variability_amplitude_vs_g}
\end{figure}

Fig. \ref{fig:variability_amplitude_vs_g} shows the variability amplitude plotted against \textit{Gaia} $G$ magnitude for the sources in our sample, colour-coded by binary type. The sample has been filtered by CSPN match certainty (reliability > 0.5) and PN angular size (radius $\geq10$\arcsec; see Sect. \ref{sect:quality} for motivation) giving 683 objects in total (see Table \ref{table:samplecount}).

Non-variable binaries (blue) include those with composite spectra and those detected (or hypothesised) based on infrared excess. Visual binaries (green) have typically been discovered through high resolution \textit{Hubble} Space Telescope imagery (and appear blended in ground-based imagery), but do not have measured radial velocity or photometric variation. Long period binaries (orange) have composite spectra and measured periods (using radial velocity variations) on the order of years. Some exhibit shorter timescale variability due to rotation. Short period or close binaries (red) generally exhibit the greatest photometric variability. These have typical periods less than a day, with only a few between 2 and 20 days. Half-filled red circles are possible short period binaries whose nature is still uncertain: the recent discoveries in TESS from \citet{aller2020tess}, objects designated as in need of further study by \citet{miszalski2011}, and Pa 5 from Kepler \citep{demarco2015kepler}. A few short period binaries had also been noted previously as visual binaries or composite spectra; in these cases the short period binarity takes precedence. The remainder of the sample (objects of unknown binarity) are shown as empty circles. Markers outlined in black are objects that have low astrometric error (RUWE < 3; see Sect. \ref{sect:quality}) and significant variability ($D>8$; see Sect. \ref{sect:newcandidates}).

The lines in the plot are lines of constant variability significance $D=0$ and $D=8$ (see Eq. \ref{eq:D}) for a source with a representative CSPN colour, that is, $G_\textup{BP}$\ --\ $G_\textup{RP}$ = 0 (this value, which is the 44th percentile of the colours of the CSPNe in the plot, is also used for sources with no colour information). The $D=8$ line is split into two different lines for 15 and 30 FoV transits, which are typical counts representing the 21st and 63rd percentile of $N_{FOV}$ for CSPNe in the sample. $D$ scales with the square root of $N_{FOV}$ ($k$ in Eq. \ref{eq:D} and preceding equations). The non-monotonic pattern in typical photometric uncertainty is evident, and is similar in shape to the magnitude error distribution in Fig. 9 of \citet{evans2018gaiaphotometry}, with peaks around $G = 11$ and $13$ due to different instrumental configurations \citep[gates and window classes;][]{carrasco2016gaiadr1photomtery} affecting calibration. Towards fainter magnitudes, where photon noise dominates, the errors increase more predictably. Though not shown in the plot, the intrinsic scatter for sources with extreme colours (e.g. very blue CSPNe) also tends to be higher, particularly at brighter magnitudes, though the overall shape of the magnitude dependence is the same.

\begin{figure}
    \centering
    \includegraphics[width=\hsize]{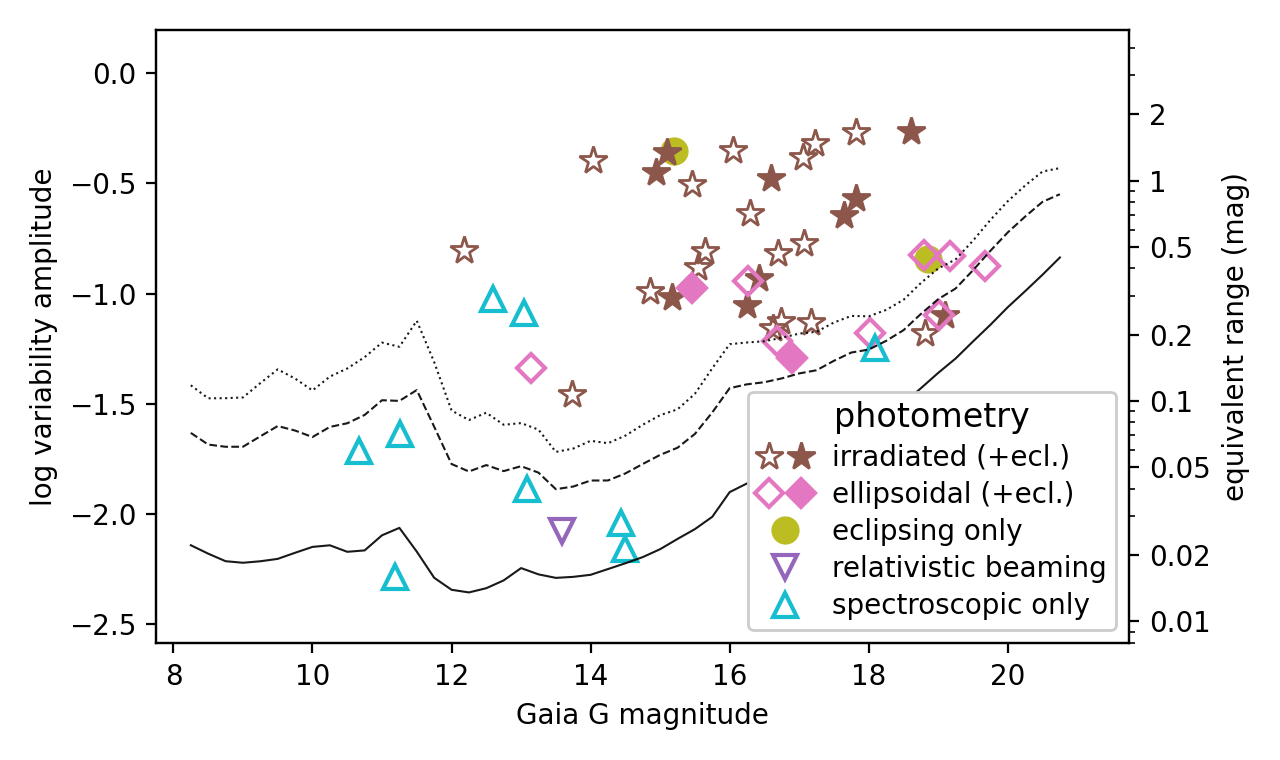}
    \caption{Same as Fig. \ref{fig:variability_amplitude_vs_g}, but for short period binaries only and with no minimum radius cut. The relaxation of the radius requirement means that not every point on this plot has a corresponding red circle in Fig. \ref{fig:variability_amplitude_vs_g}. Different marker shapes indicate different means of binarity detection, with filled markers being eclipsing (possibly in addition to another type of photometric variation).
    }
    \label{fig:variability_amplitude_short}
\end{figure}

In Fig. \ref{fig:variability_amplitude_short} we further break down known close binary systems by the type of variability. The criteria for inclusion are loosened relative to Fig. \ref{fig:variability_amplitude_vs_g} in order to increase the sample size (see counts in Table \ref{table:samplecount}). The greatest variability is evident in irradiated binaries. Ellipsoidal modulation generally produces a smaller effect, while most binaries that have been detected solely through radial velocity variation (spectroscopic binaries, shown as upward pointing triangles) have no significant photometric variability in \textit{Gaia}, being e.g. double degenerate systems such as Fg 1 \citep{boffin2012fg1} and NGC 2392 \citep{miszalski2019ngc2392}. Of the systems with low \textit{Gaia} variability, only AMU 1 has been previously identified as a photometric variable, with Kepler photometry revealing low-amplitude modulation due to relativistic Doppler boosting \citep{demarco2015kepler}.

These figures demonstrate the power of the scatter in the \textit{Gaia} mean photometry. Comparing Fig. \ref{fig:variability_amplitude_vs_g} and Fig. \ref{fig:variability_amplitude_short} we can see that all short period binaries in the sample of extended PN with CSPNe brighter than $G = 19$ that have ground-based detections of photometric variability (i.e. that are not solely spectroscopic binaries) also have significant variability in \textit{Gaia}, with variability amplitudes corresponding to peak-to-peak amplitudes of at least 0.1 mag in the $G$ band (around the limit of what is detectable from the ground). The only confidently known extended binary fainter than $G = 19$, MPA J1759-3007, is visibly separated from the trend, but has a low number of FoV transits and narrowly fails to meet the $D > 8$ significance cut.

\begin{table}
\caption{\label{table:samplecount}Sizes of samples discussed in the text.}
\centering
\begin{tabular}{lcccc}
\hline\hline
Sample&\multicolumn{4}{c}{Binary type}\\
&Unknown&Close\tablefootmark{a}&Uncertain&Other\\
\hline
all & 3427 & 60 & 12 & 44\\
\textit{Gaia} match\tablefootmark{b} & 1561 & 50\tablefootmark{c} & 10 & 40\\
radius $\geq$ 10\arcsec & 1320 & 40 & 10 & 42\\
\hline
\multicolumn{5}{c}{Extended (\textit{Gaia} match \& radius $\geq$ 10\arcsec)}\\
\hline
extended\tablefootmark{d} & 602 & 35 & 8 & 38\\
$D$ > 8 & 71 & 27 & 2 & 10\\
RUWE < 3 & 583 & 34 & 8 & 33\\
\hline
\multicolumn{5}{c}{Variable (extended \& $D$ > 8 \& RUWE < 3)}\\
\hline
variable & 58\tablefootmark{e} & 27 & 2 & 5\\
% $G < 18$ & 46 & 25 & 2 & 5\\
\hline
\end{tabular}
\tablefoot{
Each section includes the conditions from the sections above.\\
\tablefoottext{a}{Short period binaries.}
\tablefoottext{b}{Has matched \textit{Gaia} DR2 source in \citet{chornay2020cspn} catalogue with reliability > 0.5.}
\tablefoottext{c}{Sample shown in Fig. \ref{fig:variability_amplitude_short}.}
\tablefoottext{d}{Sample shown in Fig. \ref{fig:variability_amplitude_vs_g}.}
\tablefoottext{e}{New binary candidates.}
}
\end{table}

\subsection{Other sources of photometric scatter} \label{sect:quality}

Without time series photometry it is difficult to distinguish between genuine variability (possibly related to binarity) and unusually high photometric scatter. Being bright, extended objects, PNe are particularly likely to cause problems for \textit{Gaia}. In this section we consider possible sources of spurious variability.

Only single stars are modelled in the \textit{Gaia} DR2 astrometric solution. The quality of that fit is indicated by the re-normalised unit-weight error (RUWE) statistic \citep{gaiaruwe}. Unresolved binary systems can introduce wobble in the location of the flux maximum, which will contribute to a poor fit and a high value of RUWE \citep{belokurov2020ruwe}, though this is most noticeable for sources near us (within a few hundred parsecs). For visible binaries separated by less than 0.4\arcsec, astrometric issues can also be caused by the two components of a binary being detected separately and later incorrectly merged into a single source \citep{gaiadr2duplicates}, in which case the \texttt{duplicated\_source} flag will be set. Such merging is also likely to manifest as spurious photometric variability if the components have different magnitudes. Partially resolved binary systems could also cause issues for \textit{Gaia}'s point/line spread function fitting, introducing a scan angle dependence to the flux and position measurements.

\begin{figure}
    \centering
    \includegraphics[width=\hsize]{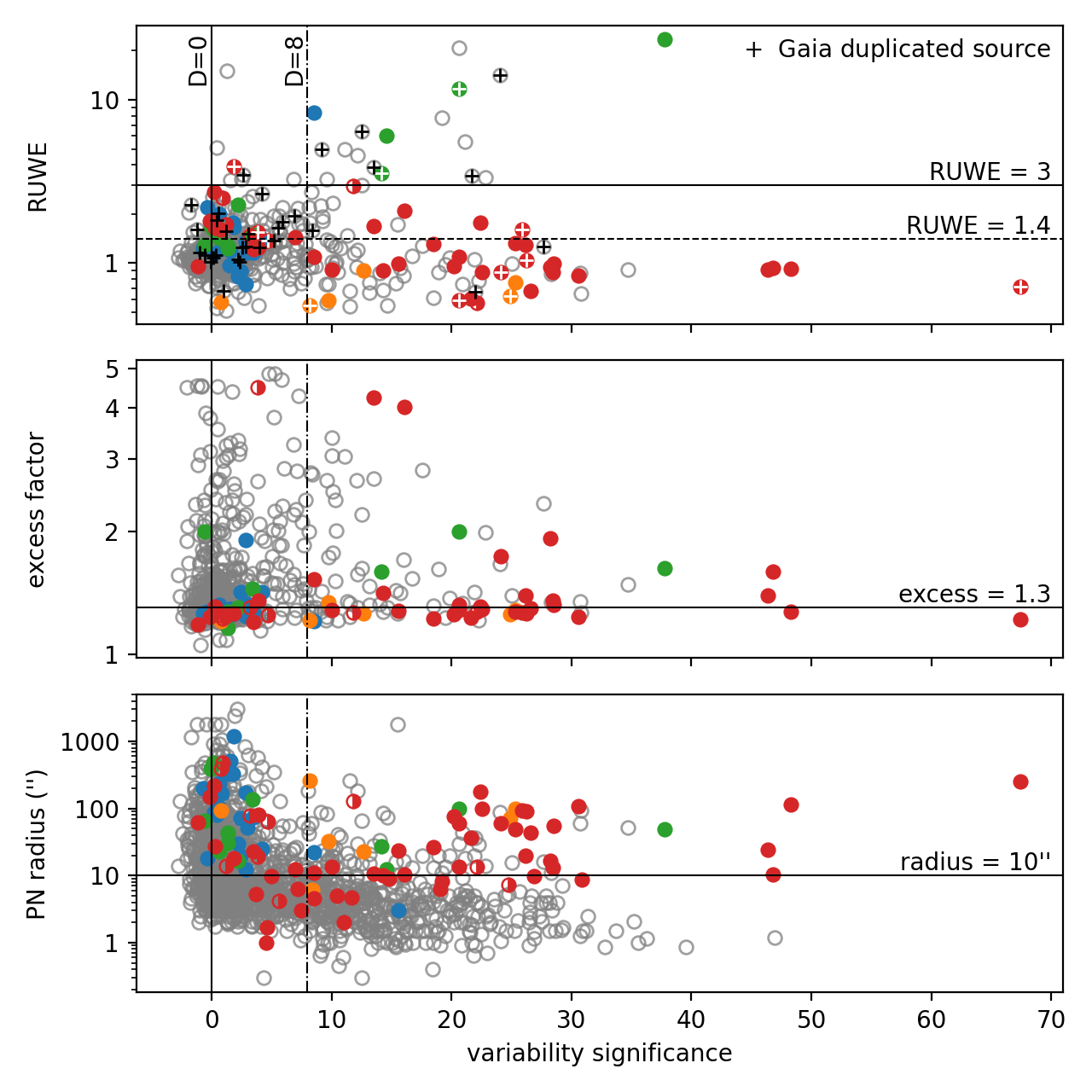}
    \caption{Quantities potentially associated with spurious variability in \textit{Gaia} photometry. Markers are different binary types as in Fig. \ref{fig:variability_amplitude_vs_g}. Vertical lines correspond to the lines of constant variability significance in Fig. \ref{fig:variability_amplitude_vs_g}. The significance of the horizontal lines is described in the text. Duplicated sources in \textit{Gaia} are indicated in the upper plot with $+$ symbols. PNe in the upper and middle plots are limited to a minimum radius of 10\arcsec\ (corresponding to the selection in Fig. \ref{fig:variability_amplitude_vs_g}), while the lower plot has no such restriction.
    }
    \label{fig:ruwe_vs_variability_amplitude}
\end{figure}

The top panel of Fig. \ref{fig:ruwe_vs_variability_amplitude} shows RUWE versus the variability amplitude significance, with the same colouring scheme as in Fig. \ref{fig:variability_amplitude_vs_g}. All known extended short period binary systems (red points) have $\textup{RUWE}<3$, with most having $\textup{RUWE}<1.4$, 1.4 being the rather conservative cutoff for well-behaved sources suggested by \citet{gaiaruwe}. Such values suggest that the photometric scatter is real rather than due to cross-matching or fit issues. Several visual binaries (green) have significant variability and high RUWE values. These pairs are typically separated by around 0.3\arcsec, a separation at which they can be resolved by \textit{Gaia} but then are merged into a single source. This merging is likely responsible for both the high photometric scatter and poor astrometric fit.

Several visual binaries also have relatively high photometric excess factors (middle panel of Fig. \ref{fig:ruwe_vs_variability_amplitude}), as the window used for $G_\textup{BP}$ and $G_\textup{RP}$ photometry captures both members of the pair. However the distinction between well-behaved sources and not is less clear than for RUWE, as other features such as crowded fields and bright nebulae can also contaminate the $G_\textup{BP}$ and $G_\textup{RP}$ photometry without necessarily introducing spurious variability in the $G$ band. The suggested excess factor threshold from \citet{evans2018gaiaphotometry} for an object with $G_\textup{BP}$\ --\ $G_\textup{RP}$ = 0 is shown for reference, though it is too conservative for our application and we do not use any cut based on excess factor.

The lower panel of Fig. \ref{fig:ruwe_vs_variability_amplitude} shows the relation between variability significance and nebula radius. There is a notable excess of high variability objects at smaller radii. This is not surprising, as these nebulae tend be brighter, and thus introduce an additional source of sky background. Compact nebulae can also cause the sorts of astrometric issues described previously. Known binaries at smaller radii do exhibit significant variability, but precisely selecting them is difficult. Thus we adopt a minimum radius of 10\arcsec\ in our selection.

\subsection{New candidate binaries} \label{sect:newcandidates}

Markers outlined in black in Fig. \ref{fig:variability_amplitude_vs_g} are objects that have significant \textit{Gaia} variability ($D>8$) and low astrometric error (RUWE < 3), in addition to meeting reliability > 0.5 and radius $\geq 10$\arcsec\ criteria common to all objects in the extended sample. We choose the criteria, in particular the threshold $D = 8$, as a balance between precision and recall, with the aim of selecting known short period binaries (red points in Fig. \ref{fig:ruwe_vs_variability_amplitude}). The 92 objects meeting these criteria represent 13\% of that sample overall, but include 27 short period binaries (77\% of those in the sample), as well as 4 long period binaries exhibiting shorter timescale variability, and 1 visual binary with a borderline RUWE value.

This leaves 58 objects that are not previously known or suspected binaries but exhibit significant variability in \textit{Gaia}. Of them, 43 have log variability amplitude $> -1.5$, suggestive of a level of variability that, if real, should be recoverable from ground-based followup. Confirming and measuring the periods of these systems, even at a moderate success rate, would dramatically increase the sample of known binary CSPNe. The list of these objects is given in Table \ref{table:catalogue}.

Such a wealth of previously undiscovered binaries is not unexpected. An unbiased ground-based survey, OGLE-III, previously found a close binary fraction of 12--21\% \citep{miszalski2009oglebinaries}. Our sample of extended PNe (radii $\geq 10$\arcsec) with well-behaved astrometry (RUWE < 3) and bright central stars ($G$ < 18) contains 318 PNe. Of these 9--11\% are known short period binaries. If we were to assume that all of the detected variables with log variability amplitude $> -1.5$ are indeed binaries (i.e. trying to recreate the limits of the OGLE survey) we would find a binary fraction of 18\%, consistent with the range from OGLE. The estimate depends on the particular thresholds chosen, and misses spectroscopic binaries, but this consistency supports the hypothesis that many of the newly discovered variables are indeed binaries.

\subsection{Nebular chemistry}

One strong indicator of PN central star binarity is a high abundance discrepancy factor \citep[ADF;][]{wesson2018adf}. This is the ratio of abundances of the same species derived from recombination lines of a photoionised nebula to those derived from collisionally excited lines. It has a median value of 2.6 for PNe. Extreme values of the ADF (ADF > 5) in PNe are associated with close binary central stars with a period of the order of a day or less. Showing that some of the variables in our sample exhibit such extreme ADFs would strengthen the link.

\begin{figure}
    \centering
    \includegraphics[width=\hsize]{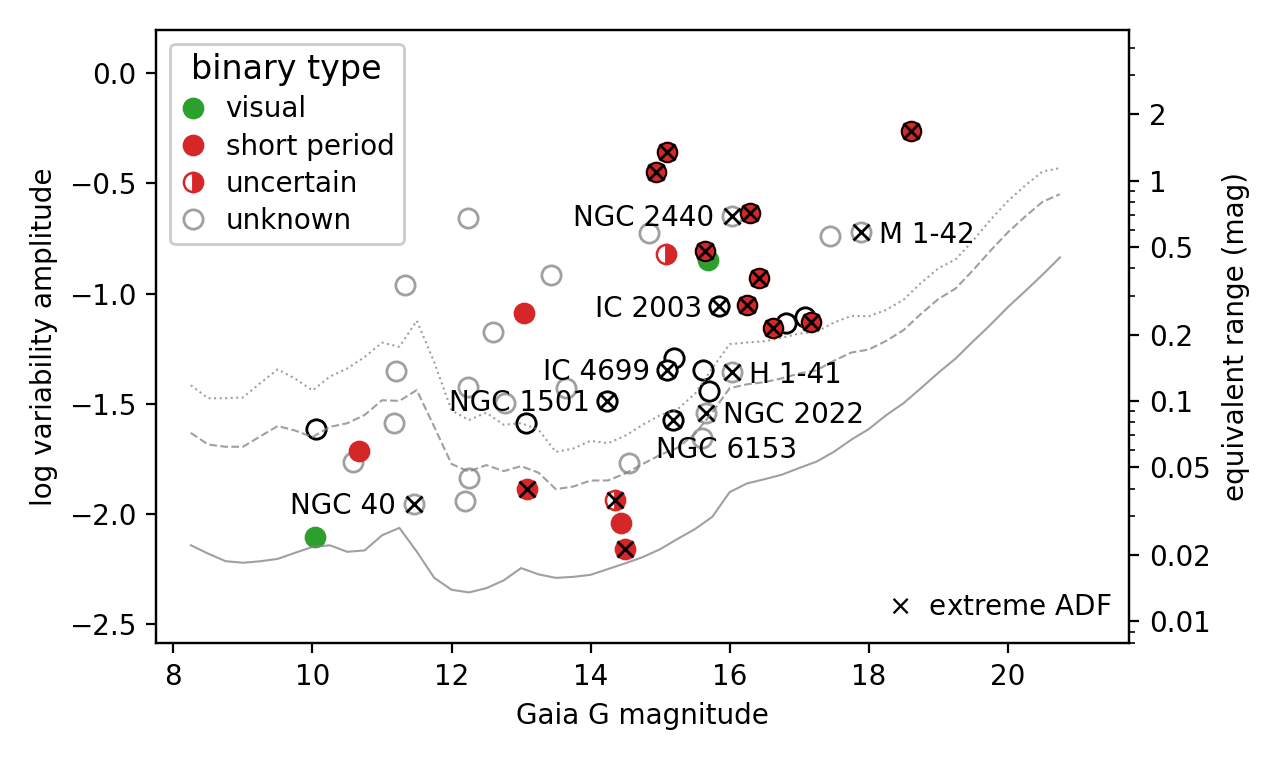}
    \caption{Extreme ADF objects (indicated with $\times$'s) on the variability versus magnitude plot. Labelled objects are PNe with extreme ADFs that are not previously known binaries. Marker colours indicate type of binarity if known, and the grey lines are the same lines of constant variability significance shown in Fig. \ref{fig:variability_amplitude_vs_g}. Likewise, markers outlined in black have the same meaning as in Fig. \ref{fig:variability_amplitude_vs_g}.}
    \label{fig:variability_amplitude_vs_adf}
\end{figure}

We join our catalogue with a compilation of literature ADF values,\footnote{\url{https://www.nebulousresearch.org/adfs/}} and again require reliability > 0.5 but relax the angular size cut to include more compact PNe (radii $\geq$ 5\arcsec). The 50 objects that passed these cuts (out of the 123 Galactic PNe with literature ADFs) are shown in Fig. \ref{fig:variability_amplitude_vs_adf}. There are 21 PNe exhibiting extreme ADFs, of which 12 are already known or suspected to harbour close binaries. 

As in Fig. \ref{fig:variability_amplitude_vs_g}, the lines indicate constant variability significance $D = 0$ and $D = 8$, and circles with black outlines are sources with both RUWE < 3 and $D > 8$. The sources above the $D = 8$ lines without black outlines have high RUWE values; the bias of the ADF sample towards compact, bright PNe, which are more amenable to spectroscopy, likely contributes to the larger proportion of astrometric issues relative to the sample in Fig. \ref{fig:variability_amplitude_vs_g}. These issues could in turn be responsible for some of the observed photometric scatter (as it seems to be for some sources in the top panel of Fig. \ref{fig:ruwe_vs_variability_amplitude}), though the observed scatter still serves as an upper bound.

With that in mind, it is reassuring that, of the extreme ADF sample, only NGC 40 does not show at least some variability. Along with NGC 40, NGC 2022 and M 1-42 have so far eluded ground-based detections of periodic variability. From Fig. \ref{fig:variability_amplitude_vs_adf} we can see that variation in the CSPN of NGC 2022 seems likely to be genuine ($D = 5.7$, RUWE = 1.4) but on the threshold of ground-based detectability, while for M 1-42 with high variability significance $D = 13.3$ but RUWE = 3.0, the situation is less clear. Of the remaining extreme ADF objects, four have significant variability and low RUWE (though most of their amplitudes are again likely to be difficult to detect from the ground), while NGC 2440 has an extreme RUWE value that makes its apparent variability inconclusive, and H 1-41 has a moderate variability significance of $D = 6.8$, which may be improved in the future with more observations.

Thus, from the \textit{Gaia} data alone, we can conclude that NGC 40 is unlikely to have strong photometric variability (though that of course does not preclude it from being a binary), but only state for other extreme ADF objects that their photometric scatters are consistent with variability caused by binarity. Light curves are needed to ascertain whether there is a periodic component to the scatter of these objects, and whether the period matches the expectation from \citet{wesson2018adf}, who found that extreme-ADF PNe have central stars with periods of less than 1.15 days.

\section{Light curves from epoch photometry}

Characterisation of variable sources in \textit{Gaia} DR2 is limited to about 550\,000 sources that were classified as \texttt{VARIABLE} by \textit{Gaia}'s own analysis \citep{holl2018variability}. Epoch photometry (flux measurements for a source at each FoV transit) is published for all of these sources, as well as a table (\texttt{vari\_time\_series\_statistics}) of summary statistics derived from that photometry. Around 3000 sources exhibiting short timescale variability additionally have frequency information published in the \texttt{vari\_short\_timescale} table.

However completeness is low: only four of the known short period binary CSPNe are classified as \texttt{VARIABLE}, despite many more being known to exhibit strong photometric variability, and only two of them have frequencies recovered by \textit{Gaia}. Additionally four other sources are classified as \texttt{VARIABLE} that are not previously known binaries. All eight of these \texttt{VARIABLE} sources were also identified in our selection as significantly variable based on their flux uncertainties. Though they represent only a fraction of such sources, their epoch photometry offers a window through which to probe the nature of their variability, as well as the assumptions that went into our analysis.

\subsection{Frequency extraction}

We performed Lomb-Scargle analysis \citep[using the implementation of][]{vanderplas2015} on the eight $G$ band light curves in order to extract periodic variability, with a grid of 10$^6$ frequencies logarithmically spaced between $10^{-1.5}$ and $10^{1.5}$ days$^{-1}$. We select the frequency at the peak with the highest power, and fit the light curves to a sinusoid
\begin{equation} \label{eq:sine}
    G(t) = A \sin(2\pi f t + \phi) + c,
\end{equation}
where $A$ and $c$ are the amplitude and offset in mags, $f$ is the frequency in days$^{-1}$, $t$ is the Barycentric Julian Date (JD) in Barycentric Coordinate Time (TCB) relative to \textup{2010-01-01T00:00:00}, and $\phi$ is the phase. We determine these parameters through least squares optimisation, allowing the frequency to vary within a small range close to the highest power frequency extracted from the periodogram.

\subsection{Results}

\begin{figure*}
    \centering
    \includegraphics[width=\hsize]{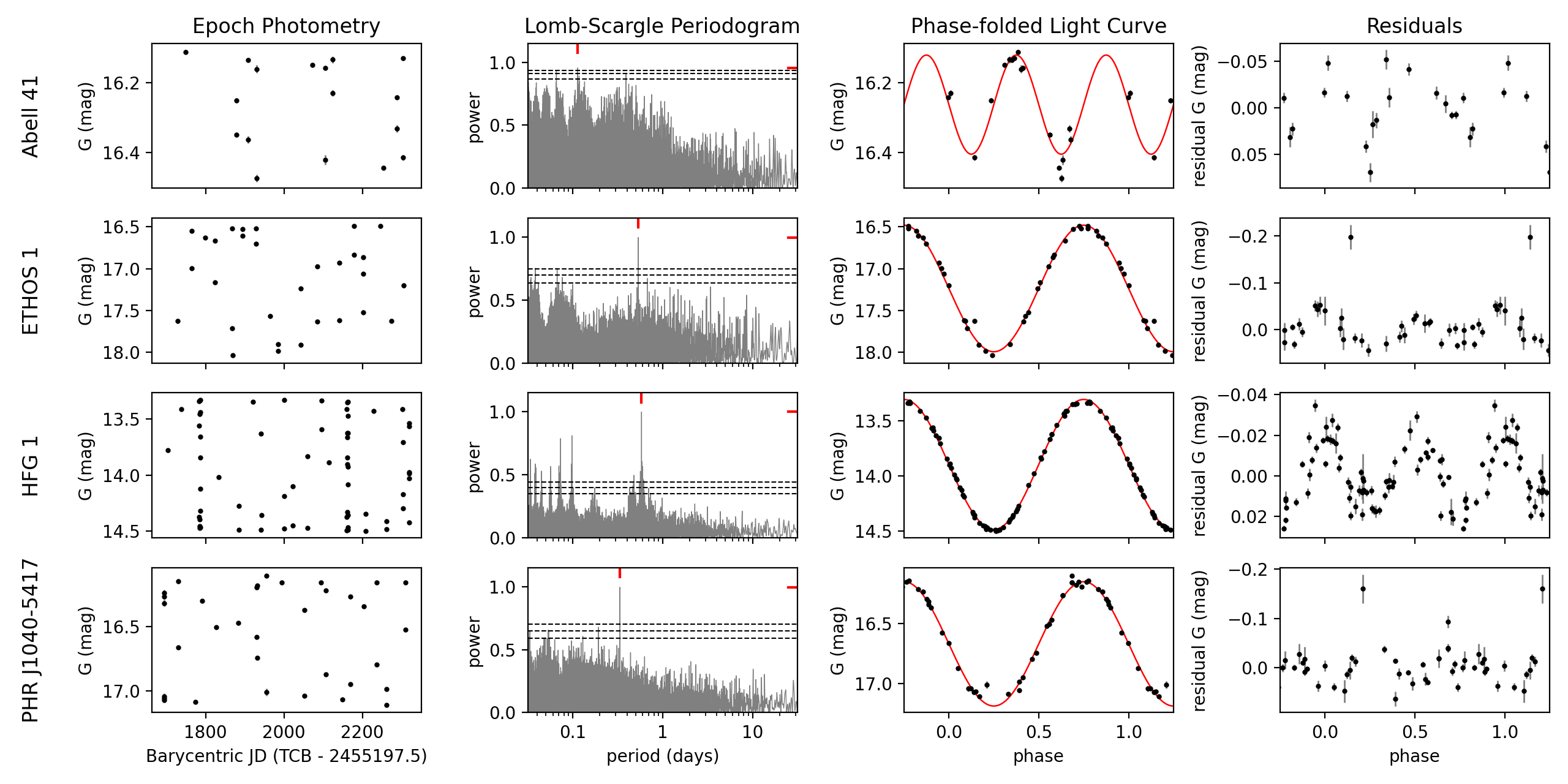}
    \caption{Epoch photometry (left), derived periodograms (left centre), phase-folded light curves (right centre) and residuals (right) for previously known binaries classified as \texttt{VARIABLE} by \textit{Gaia}. Red ticks on the periodograms indicate the highest peak, and horizontal dashed lines are false alarm levels at (from low to high) 10\%, 1\%, and 0.1\%. Powers are normalised relative to the residuals of a constant model, and lie between 0 (no improvement over constant model) and 1 (periodic model has zero residuals). Error bars are $2\sigma$ errors for visibility. An additional half phase is shown for the light curves and residuals. The light curve and residuals for Abell 41 are folded at double the derived period (see text).}
    \label{fig:light_curves_binaries}
\end{figure*}

\begin{figure*}
    \centering
    \includegraphics[width=\hsize]{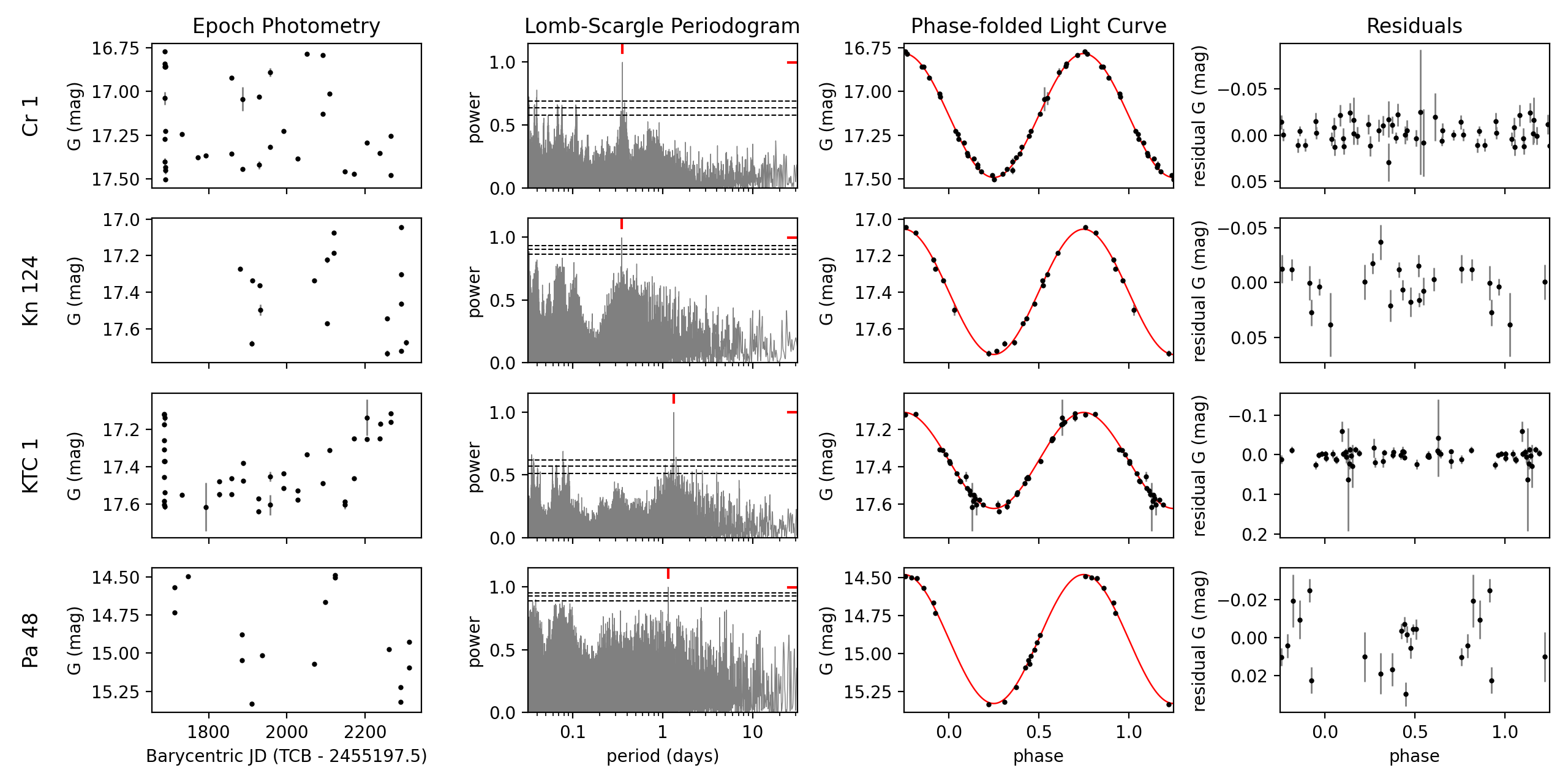}
    \caption{Same as Fig. \ref{fig:light_curves_binaries}, but for new candidate binaries.}
    \label{fig:light_curves_new}
\end{figure*}

Table \ref{tbl:1} contains the parameters of the best fits. Derived phase-folded light curves and residuals are shown in Fig. \ref{fig:light_curves_binaries} for known binaries and Fig. \ref{fig:light_curves_new} for new candidates. All eight \texttt{VARIABLE} sources have good fits, and three out of the four derived periods for known binary systems also match literature values.

\begin{table}
\caption{\label{tbl:1}Parameters of the best sinusoidal fits of \textit{Gaia} epoch photometry for known binary systems and newly discovered variables.}
\centering
\begin{tabular}{lllll}
\hline\hline
PN & $A$ (mag) & $T$ (days) & $\phi$ & $c$\\
\hline
\multicolumn{5}{c}{Known binaries}\\
\hline
Abell 41 & 0.142 & 0.1132\tablefootmark{a} & 3.833 & 16.261 \\
ETHOS 1 & 0.754 & 0.5351\tablefootmark{b,c} & 1.668 & 17.234 \\ % 0.5351263
HFG 1 & 0.589 & 0.5817\tablefootmark{d} & 2.607 & 13.896 \\ %+ 0.5817(2)
PHR J1040-5417 & 0.516 & 0.3347\tablefootmark{c,e} & 2.154 & 16.669 \\
\hline
\multicolumn{5}{c}{New binary candidates}\\
\hline
Cr 1 & 0.355 & 0.3569 & 3.457 & 17.137 \\
Kn 124 & 0.336 & 0.3492 & 5.628 & 17.397 \\
KTC 1 & 0.257 & 1.3290 & 4.775 & 17.366 \\
Pa 48 & 0.427 & 1.1511 & 4.792 & 14.905 \\
\hline
\end{tabular}
\tablefoot{See Eq. \ref{eq:sine} for definitions. $T\equiv1/f$.
\tablefoottext{a}{Matches one possible model from \citet{bruch2001abell41}; later work by \citet{jones2010abell41} determined that the period is in fact twice this value.}
\tablefoottext{b}{Matches \citet{munday2020ethos1}.}
\tablefoottext{c}{Also matches period in \textit{Gaia} \texttt{vari\_short\_timescale} table.}
\tablefoottext{d}{Matches \citet{exter2005hfg1}.}
\tablefoottext{e}{Matches \citet{hillwig2017phrj1040}.}
}
\end{table}

One exception is Abell 41, where our derived period is half that in literature \citep{jones2010abell41}. The variability of Abell 41 is believed to be due to ellipsoidal modulation and partial eclipses, so it has two minima per orbit, and the shape of its folded light curve is different from the others in the sample and less well described by a sinusoid. We have plotted the folded light curve for Abell 41 at double our derived period to reflect this, though the difference e.g. in the depths of the minima is difficult to see with \textit{Gaia}'s sparse sampling \citep[cf. Fig. 2 of ][]{bruch2001abell41}.

The known irradiated binary PHR J1040-5417 has deep primary eclipses \citep{hillwig2017phrj1040} that are not visible in the epoch photometry; instead there is a gap at the light curve minimum. The eclipse is deep enough that it would be close to the detection limit of \textit{Gaia} DR2; however comparison with the \textit{Gaia} scanning law shows that this gap also exists in the sampling.

The four newly discovered variable CSPNe are those of Cr 1,  KTC 1, Kn 124, and Pa 48. All of these PNe are relatively recent discoveries; only Cr 1 and KTC 1 \citep{jacobydsssearch} have spectroscopic confirmation of their PN status. Kn 124 and Pa 48 \citep{kronberger2016prc15} do not and are only listed as possible and likely PNe in the HASH PN catalogue. None have published narrowband imagery or morphological information.

The $G$ band light curves of these newly discovered variables are well described by sinusoidal fits. The shape, amplitude, and periods of the curves are consistent with irradiation effects, in which one side of a cool companion is heated by the CSPN and contributes to the flux of the system most when it is directly behind the CSPN; i.e. the hot side is facing the observer. With irradiation effects there is one flux maximum per orbit, so the period of the variation is the period of the binary system. None of the systems appear to be eclipsing, and the uncertainty in $G_\textup{BP}$ and $G_\textup{RP}$ is too large to measure any colour effects. We conclude that they are likely binaries, though constraining the parameters of the systems is not possible without additional data and is beyond the scope of this work.

\subsection{Relation to mean photometry}

The additional data in the epoch photometry tables help to validate some of our assumptions in Sect. \ref{sect:methods} albeit with a very small sample. For the eight \texttt{VARIABLE} sources in the \texttt{vari\_time\_series\_statistics} table, the mean of \texttt{phot\_g\_n\_obs} $/$ \texttt{num\_selected\_g\_fov} (i.e. $N_{CCD} / N_{FOV}$) is $8.89 \pm 0.29$, consistent with our choice of $8.86$. The mean ratio of $G$ range to $A_{var,G}$ is $3.15 \pm 0.10$, slightly higher than but consistent with the value in Eq. \ref{eq:a_sine} (though Eq. \ref{eq:a_sine} applies only in the limit of uniform dense sampling).

\begin{figure}
    \centering
    \includegraphics[width=\hsize]{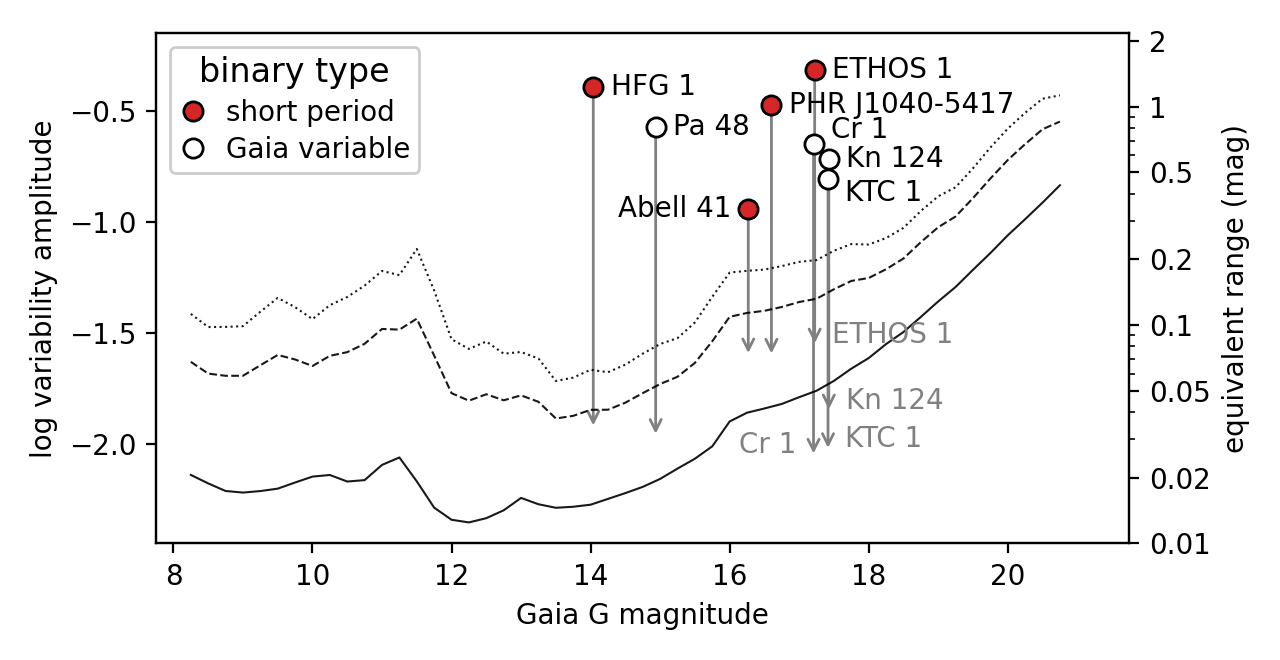}
    \caption{Initial variability (points) and residual variability (arrow tips) of \textit{Gaia} \texttt{VARIABLE} sources after subtracting sinusoidal fits from the epoch photometry. The black lines are the same lines of constant variability significance  shown in Fig. \ref{fig:variability_amplitude_vs_g}.}
    \label{fig:residual_variability_amplitude}
\end{figure}

Subtracting the best fit sinusoids removes most of the periodic variability, though HFG 1 in particular shows some in its residuals (Fig. \ref{fig:light_curves_binaries}, right panels) that would be easily accounted for by adding a second higher-frequency term to its fit. The amplitude of the remaining variability (derived from the weighted variance of the residuals of the sinusoidal fits) is shown in Fig. \ref{fig:residual_variability_amplitude}. The significance of most of the residuals is low. The low residual variability of Cr 1 and KTC 1 compared to typical CSPNe is due to their relatively red central stars, which have a lower intrinsic measurement uncertainty.

\section{Conclusions}

\textit{Gaia} shows tremendous promise for the study of CSPNe. Though \textit{Gaia}'s own variability processing is quite limited, combining the scatter in the photometry with the notion of variability significance introduced in this work allows us to retrieve a large fraction of the binary systems that have been discovered from painstaking ground-based observations over the past decades. The newly discovered likely variables offer a promising set of targets for ground-based photometric followups, building towards a statistically significant sample of post-AGB, post-CE stars. Many of their nebulae have chemistry or morphology typically associated with binarity. Selecting targets on variability, rather than expectations of binarity based on other features, will help us study how these features relate to binarity in a less biased manner, in addition to increasing the expected success rate.

With available \textit{Gaia} epoch photometry we have shown that we can recover periods for four known binary CSPNe, and confirmed that four of the previously unknown variable sources have light curves consistent with binarity. This lends weight to our hypothesis that many of the variables that we have identified from their mean photometry are indeed binaries.

Completeness and accuracy will improve with future \textit{Gaia} data releases, with more transits increasing the statistical significance of any observed excess variability. Improved photometric processing and handling of non-single systems will reduce spurious sources of photometric scatter, and ultimately epoch photometry will enable us to confirm new binary candidates from the \textit{Gaia} data alone.

\begin{acknowledgements}
We thank the anonymous referee for their comments, which have helped improve this paper.

We also thank Todd Hillwig for providing a reference light curve for PHR J1040-5417.

This research has made use of data from the European Space Agency (ESA) mission \textit{Gaia} (\url{https://www.cosmos.esa.int/gaia}) processed by the {\it Gaia} Data Processing and Analysis Consortium (DPAC, \url{https://www.cosmos.esa.int/web/gaia/dpac/consortium}). Funding for the DPAC has been provided by national institutions, in particular the institutions participating in the {\it Gaia} Multilateral Agreement.
This research has also made use of the HASH PN database (\url{http://hashpn.space}) and of Astropy (\url{http://www.astropy.org}), a community-developed core Python package for Astronomy \citep{astropy2013, astropy2018}.

NC is supported through the Cancer Research UK grant A24042.

DJ acknowledges support from the State Research Agency (AEI) of the Spanish Ministry of Science, Innovation and Universities (MCIU) and the European Regional Development Fund (FEDER) under grant AYA2017-83383-P.  DJ also acknowledges support under grant P/308614 financed by funds transferred from the Spanish Ministry of Science, Innovation and Universities, charged to the General State Budgets and with funds transferred from the General Budgets of the Autonomous Community of the Canary Islands by the Ministry of Economy, Industry, Trade and Knowledge.  DJ also acknowledges support from the Erasmus+ programme of the European Union under grant number 2020-1-CZ01-KA203-078200.
\end{acknowledgements}

\bibliographystyle{aa} % style aa.bst
\bibliography{bib.bib} % your references Yourfile.bib

\begin{appendix}
\section{Table of new candidate binaries}

\begin{table*}
\scriptsize
\centering
\caption{\label{table:catalogue}New candidate binaries.}
\setlength\tabcolsep{5pt} % default value: 6pt
\begin{tabular}{cccccccccccc}
\hline \hline
PN name & radius & RA & Dec & \textit{Gaia} DR2 source ID & reliability\tablefootmark{a} & $G$ & $G_\textup{BP}$--$G_\textup{RP}$ & RUWE & $\log A_{var,G}$\tablefootmark{b} & range $G$\tablefootmark{c} & $D$\tablefootmark{d}\\
 & (\arcsec) & (deg) & (deg) &  &  & (mag) & (mag) &  &  & (mag) &  \\
\hline
Pa 30 & 85.5 & 13.2967 & 67.5007 & 526287189172936320 & 1.00 & 15.40 & 0.84 & 0.68 & -1.57 & 0.08 & 14.3 \\
WeSb 1 & 92.5 & 15.2254 & 55.0667 & 423384961080344960 & 1.00 & 14.81 & 1.19 & 0.65 & -0.98 & 0.32 & 30.8 \\
NGC 1501 & 28.5 & 61.7475 & 60.9206 & 473712872456844544 & 1.00 & 14.24 & 0.57 & 0.78 & -1.49 & 0.10 & 22.3 \\
Kn 51 & 42.0 & 66.3619 & 35.1022 & 176269718435866112 & 1.00 & 18.24 & 1.74 & 0.95 & -1.20 & 0.19 & 8.1 \\
Kn 133 & 15.5 & 73.6403 & 28.8247 & 155075188702503424 & 1.00 & 18.96 & 0.95 & 0.99 & -0.44 & 1.13 & 25.1 \\
IPHAS J052015.87+314938.8 & 20.0 & 80.0661 & 31.8274 & 180731845860251264 & 0.99 & 19.24 & 1.49 & 1.07 & -0.32 & 1.46 & 19.8 \\
Abell 10 & 18.6 & 82.9404 & 6.9339 & 3333922699432278656 & 1.00 & 18.65 & 0.17 & 1.26 & -0.81 & 0.47 & 10.3 \\
PHR J0633-0135 & 30.0 & 99.5672 & -1.5524 & 3106907671813244032 & 0.97 & 16.43 & 1.85 & 1.20 & -1.18 & 0.20 & 12.2 \\
HaWe 8 & 52.5 & 100.0404 & 21.4171 & 3378885131503804544 & 1.00 & 18.92 & 0.60 & 1.14 & -0.89 & 0.40 & 8.0 \\
M 3-4 & 16.5 & 118.7975 & -23.6368 & 5710725616423348352 & 1.00 & 15.22 & 1.46 & 0.67 & -1.72 & 0.06 & 11.5 \\
BRAN 229 & 73.5 & 136.4210 & -47.9014 & 5327072405662519808 & 0.64 & 12.16 & 1.55 & 0.55 & -1.59 & 0.08 & 14.7 \\
IC 2448 & 11.0 & 136.7763 & -69.9418 & 5222772389050179840 & 1.00 & 14.01 & -0.62 & 2.20 & -1.42 & 0.12 & 8.1 \\
PHR J0928-4936 & 43.5 & 142.1707 & -49.6130 & 5313911079687293440 & 1.00 & 16.94 & 1.66 & 0.74 & -1.11 & 0.24 & 20.9 \\
WPS 54 & 1800.0 & 147.8583 & 53.1585 & 1020641700210987520 & 1.00 & 15.27 & -0.62 & 1.73 & -1.16 & 0.21 & 15.5 \\
PHR J1007-5304 & 20.0 & 151.8008 & -53.0790 & 5356821582547892352 & 0.95 & 18.65 & 0.91 & 1.05 & -0.71 & 0.60 & 21.9 \\
PHR J1007-6124 & 13.0 & 151.9639 & -61.4067 & 5256301476452488320 & 0.95 & 19.78 & 0.95 & 1.10 & -0.29 & 1.59 & 16.7 \\
NGC 3195 & 19.8 & 152.3368 & -80.8584 & 5198687896083997312 & 1.00 & 17.53 & 0.11 & 1.30 & -1.23 & 0.18 & 10.0 \\
PHR J1134-5243 & 21.0 & 173.6606 & -52.7256 & 5345597458614614400 & 0.99 & 13.00 & 0.17 & 1.39 & -1.82 & 0.05 & 9.6 \\
Hen 2-70 & 17.3 & 173.7956 & -60.2836 & 5335879596943573888 & 0.98 & 15.66 & 1.94 & 0.75 & -1.67 & 0.07 & 9.8 \\
Th 2-A & 13.7 & 200.6414 & -63.3506 & 5865192156852977536 & 0.99 & 17.04 & $\ldots$ & 1.92 & -1.35 & 0.14 & 9.2 \\
IC 4406 & 23.2 & 215.6089 & -44.1506 & 6102472263541814272 & 0.99 & 16.81 & $\ldots$ & 1.80 & -1.13 & 0.23 & 9.3 \\
PHR J1424-5138 & 59.5 & 216.1348 & -51.6442 & 5898716231900175616 & 0.95 & 13.00 & 0.05 & 2.74 & -1.65 & 0.07 & 8.3 \\
Hen 2-111 & 14.7 & 218.3270 & -60.8268 & 5878442989838460416 & 0.96 & 16.45 & 1.14 & 0.94 & -1.66 & 0.07 & 10.1 \\
Pa 63 & 41.0 & 223.5388 & -56.8047 & 5881036394197872384 & 0.99 & 11.34 & 0.73 & 0.88 & -1.45 & 0.11 & 10.5 \\
BMP J1533-5319 & 16.5 & 233.3361 & -53.3169 & 5888601549601893632 & 0.82 & 16.81 & 2.96 & 0.84 & -1.20 & 0.20 & 16.0 \\
MPA J1602-5543 & 16.0 & 240.5466 & -55.7252 & 5836526376829511808 & 0.96 & 15.89 & 2.19 & 0.75 & -1.44 & 0.11 & 15.4 \\
Sand 3 & 180.0 & 241.6186 & -35.7535 & 6010805807350513920 & 1.00 & 13.91 & 0.21 & 1.06 & -2.00 & 0.03 & 8.0 \\
NGC 6072 & 37.1 & 243.2432 & -36.2298 & 6022549519257057280 & 1.00 & 18.27 & $\ldots$ & 1.24 & -1.23 & 0.18 & 10.7 \\
Mz 3 & 23.9 & 244.3058 & -51.9863 & 5934701559547878144 & 0.98 & 13.18 & 1.82 & 2.34 & -1.90 & 0.04 & 10.1 \\
BMP J1622-5144 & 26.0 & 245.6418 & -51.7488 & 5934018728455510528 & 0.90 & 18.62 & 1.32 & 0.99 & -1.16 & 0.21 & 10.3 \\
NGC 6153 & 13.5 & 247.8774 & -40.2535 & 6017034570775817984 & 0.89 & 15.18 & 0.64 & 1.02 & -1.57 & 0.08 & 10.4 \\
PHR J1641-5302 & 10.2 & 250.2683 & -53.0402 & 5930814309267559552 & 1.00 & 16.51 & 0.79 & 0.93 & -1.26 & 0.17 & 15.7 \\
K 2-16 & 13.3 & 251.2044 & -28.0680 & 6033785806546839936 & 1.00 & 12.09 & 0.46 & 0.98 & -1.41 & 0.12 & 19.5 \\
PHR J1700-5047 & 10.5 & 255.1573 & -50.7900 & 5936569462322610688 & 0.95 & 17.32 & 0.20 & 1.11 & -1.19 & 0.20 & 8.3 \\
K 2-17 & 19.6 & 257.3994 & -52.2173 & 5924383368793138560 & 0.95 & 17.18 & 0.54 & 0.94 & -1.43 & 0.11 & 8.3 \\
Kn 124 & 20.0 & 258.5934 & -28.4576 & 4107710952444522496 & 0.93 & 17.43 & 0.57 & 0.88 & -0.72 & 0.59 & 18.9 \\
PTB 26 & 15.0 & 262.3053 & -16.7955 & 4124562102059801216 & 1.00 & 16.64 & $\ldots$ & 1.14 & -0.99 & 0.31 & 11.8 \\
NGC 6369 & 15.0 & 262.3352 & -23.7597 & 4111368477921050368 & 0.78 & 15.62 & 1.60 & 0.76 & -1.34 & 0.14 & 13.2 \\
PHR J1752-3330 & 13.0 & 268.1217 & -33.5012 & 4043198031104721664 & 0.98 & 14.86 & 0.93 & 0.61 & -1.07 & 0.26 & 18.5 \\
PHR J1757-1649 & 21.0 & 269.4151 & -16.8221 & 4144686669576773888 & 0.99 & 15.66 & 1.21 & 0.66 & -1.27 & 0.17 & 13.1 \\
PTB 19 & 10.0 & 269.6085 & -14.4232 & 4148581650775253120 & 0.81 & 18.39 & 1.28 & 0.94 & -0.90 & 0.39 & 12.1 \\
SB 40 & 10.2 & 270.7322 & -37.1373 & 4037160548451676416 & 0.53 & 17.48 & $\ldots$ & 1.76 & -1.17 & 0.21 & 8.2 \\
K 6-41 & 11.8 & 272.8023 & -29.3667 & 4050459652633253376 & 0.63 & 18.35 & 1.11 & 1.58 & -1.12 & 0.24 & 8.4 \\
Abell 44 & 33.5 & 277.2939 & -16.5348 & 4097048709015654784 & 0.77 & 16.98 & 0.80 & 0.87 & -1.05 & 0.28 & 14.0 \\
We 3-1 & 87.5 & 278.5111 & 14.8221 & 4509807233807831936 & 1.00 & 15.20 & 0.96 & 2.24 & -1.74 & 0.06 & 9.1 \\
Pa 120 & 11.5 & 283.4311 & -14.3328 & 4101903087213994496 & 1.00 & 14.30 & 0.44 & 0.74 & -1.80 & 0.05 & 9.5 \\
IPHASX J191716.4+033447 & 16.0 & 289.3183 & 3.5798 & 4292267621344388864 & 0.84 & 14.31 & 1.47 & 0.57 & -1.87 & 0.04 & 9.6 \\
NGC 6804 & 29.1 & 292.8964 & 9.2253 & 4296362443149857920 & 1.00 & 14.00 & 0.49 & 1.35 & -1.92 & 0.04 & 9.9 \\
Kn 15 & 13.2 & 295.1681 & 29.5029 & 2031657180462828288 & 1.00 & 17.69 & 0.33 & 0.96 & -1.30 & 0.15 & 8.7 \\
Pa 161 & 262.5 & 295.8691 & -13.7498 & 4183333506776770688 & 1.00 & 13.33 & 1.07 & 0.55 & -1.59 & 0.08 & 11.6 \\
Pa 164 & 52.5 & 299.3468 & 23.8801 & 1834171384397003264 & 0.71 & 18.16 & 1.30 & 0.91 & -0.30 & 1.53 & 34.7 \\
Hen 1-5 & 16.0 & 302.9836 & 20.3345 & 1828750899461025536 & 0.96 & 17.29 & 3.08 & 1.27 & 0.07 & 3.60 & 27.7 \\
NGC 6905 & 21.6 & 305.5958 & 20.1045 & 1816547660416810880 & 1.00 & 14.55 & -0.28 & 1.47 & -1.41 & 0.12 & 11.3 \\
Pa 27 & 36.0 & 312.2432 & 32.3041 & 1859955657931121536 & 1.00 & 12.28 & 1.23 & 0.67 & -1.42 & 0.12 & 22.0 \\
NGC 7026 & 19.5 & 316.5774 & 47.8519 & 2165238733564304768 & 1.00 & 15.20 & 0.57 & 1.27 & -1.29 & 0.16 & 17.6 \\
KTC 1 & 11.0 & 322.0457 & 58.8764 & 2179544655458448512 & 0.92 & 17.41 & 1.23 & 0.86 & -0.81 & 0.48 & 28.3 \\
Cr 1 & 60.0 & 327.2987 & 57.4555 & 2202260634408052224 & 1.00 & 17.21 & 1.40 & 0.87 & -0.65 & 0.69 & 30.7 \\
NGC 7354 & 16.5 & 340.0828 & 61.2857 & 2201080755349789568 & 1.00 & 17.13 & 0.97 & 1.02 & -1.45 & 0.11 & 10.0 \\
\hline
\end{tabular}
\tablefoot{
\tablefoottext{a}{Certainty of CSPN identification (between 0 and 1) from \citet{chornay2020cspn}.}
\tablefoottext{b}{Base-10 logarithm of the variability amplitude; see Eq. \ref{eq:a_varG}.}
\tablefoottext{c}{Estimated range (peak to trough amplitude) of variation if sinusoidal; see Eq. \ref{eq:a_sine}}.
\tablefoottext{d}{Variability amplitude significance; see Eq. \ref{eq:D}.}
}
\end{table*}

\end{appendix}

\end{document}